\documentclass[12pt]{article}
\newcommand{\be} {\begin{equation}}
\newcommand{\ee} {\end{equation}}
\newcommand{\bdm} {\begin{displaymath}}
\newcommand{\edm} {\end{displaymath}}
\newcommand{\bc} {\begin{center}}
\newcommand{\ec} {\end{center}}
\newcommand{\beqa} {\begin{eqnarray}}
\newcommand{\eeqa} {\end{eqnarray}}

\newcommand{\ra} {\rightarrow}
\begin{document}
\title{\centering LIGHT VECTOR MESON SPECTROSCOPY}
\author{A. Donnachie, University of Manchester, England} 
\maketitle
\begin{abstract}
The current situation for vector meson spectroscopy is outlined, and it
is shown that the data are inconsistent with the generally-accepted model
for meson decay. A possible resolution in terms of exotic (hybrid) mesons 
is given. Although this hypothesis resolves some of the issues, fresh 
theoretical questions are raised. It is argued that high-precision $e^+e^-$
annihilation data provide an excellent laboratory for studying many aspects
of nonperturbative QCD. 
\end{abstract}

\section{THE PROBLEM}

It is now 15 years since it was first suggested \cite{EO86,DM87} that the 
$\rho'(1600)$, as it was then known, is in fact a composite structure,
consisting of at least two states: the $\rho(1450)$ and $\rho(1700)$.
Their existence, and that of their isoscalar counterparts, the 
$\omega(1420)$ and $\omega(1650)$, and of an associated hidden-strangeness 
state, the $\phi(1680)$, is now well established \cite{PDG}. 
The key data in determining the existence of the two isovector states were 
$e^+e^- \ra \pi^+\pi^-$ \cite{r1} and $e^+e^- \ra \omega\pi$ \cite{r2}. 
These original data sets have subsequently been augmented by data on the 
corresponding charged channels in $\tau$ decay \cite{r3,r4,r4a,r4b}, to 
which they are related by CVC. These new data confirm the earlier conclusions.
The data on $e^+e^- \ra  \pi^+\pi^-\pi^+\pi^-$ \cite{r5,r5a} and $e^+e^- \ra 
\pi^+\pi^-\pi^0\pi^0$ \cite{r5,r6} (excluding $\omega\pi$) and the 
corresponding charged channels in $\tau$ decay \cite{r4,r4a} are consistent 
with the two-resonance interpretation \cite{r7,r8}, although they do not 
provide such good discrimination. Itwas also found that the $e^+e^- \ra 
\eta\pi^+\pi^-$ cross section is better fitted with two interfering 
resonances than with a single state \cite{r9}.  Independent evidence for 
two $J^P = 1^-$ states is provided in a high statistics study of the
$\eta\pi\pi$ system in $\pi^- p$ charge exchange \cite{r10}. Decisive 
evidence for both the $\rho(1450)$ and $\rho(1700)$ in their $2\pi$ and
$4\pi$ decays has come from the study of $\bar{p} p$ and $\bar{p} n$ 
annihilation \cite{r11}. The data initially available for the study of 
the $\omega(1420)$ and $\omega(1600)$ were $e^+e^- \ra \pi^+\pi^-\pi^0$
\cite{r5a,r12} (which is dominated by $\rho\pi$) and $e^+e^- \ra 
\omega\pi^+\pi^-$ \cite{r12}. The latter cross section shows a clear peak 
which is apparently dominated by the $\omega(1600)$. The former cross 
section is more sensitive to the $\omega(1420)$. More recent and more
precise data \cite{SND} on $e^+e^- \ra \pi^+\pi^-\pi^0$ below 1.4 GeV
confirm \cite{r8} the $\omega(1420)$. 

However, although there is general consensus on the existence of these 
states, there is considerable disparity on the masses and widths
of these resonances. Information on the vector states comes principally
from $e^+e^-$ annihilation and $\tau$ decay, but there are problems with 
much of the data:

$\bullet$ inconsistencies, even in recent high-statistics data

$\bullet$ restricted energy ranges, e.g. Novosibirsk and CLEO

$\bullet$ poor statistics in some channels and missing channels

$\bullet$ inadequate knowledge of multiparticle final states

$\bullet$  $\sum \sigma_{\rm exclusive} > \sigma_{\rm inclusive}$

In fact the only channel with really consistent data sets over a wide energy
range is the $\pi\pi$ channel, although even that runs out of statistics at 
the upper end of the energy range. The comparatively low statistics of the
older data, the restricted energy range of the newer data with higher 
statistics, and persisting inconsistencies in data from different experiments
rule out precision fits.

There are also theoretical uncertainties which affect the analysis of 
$e^+e^-$ annihilation and $\tau$ decay, and which present data are 
insufficiently precise to resolve. Firstly there is the ``tail-of-the-$\rho$''
problem. In some channels, most notably $\pi\pi$ and $\pi\omega$, there is 
strong interference between the high-energy tail of the $\rho$ and the 
higher-mass resonances. The magnitude and shape of this tail are not known 
with any precision. They can only be specified in models and strictly should 
be part of the parametrisation. Different models yield different results for 
the masses and widths of the resonances. A related problem is the question of 
the relative phases. These can be specified in simple models, but we know 
that these models are not precise and leaving the phases as free parameters 
has a major effect. 

Thus only the qualitative features emerge, and apparent precision is in 
reality the result of implicit, or explicit, theoretical assumptions.

Despite these difficulties, the existence of the higher-mass states is not 
in doubt, and a natural explanation for them  is that they are the first 
radial, $2^3S_1$, and first orbital, $1^3D_1$, excitations of the $\rho$ 
and $\omega$ and the first radial excitation of the $\phi$, as the 
generally-accepted masses \cite{PDG} are close to those predicted by the 
quark model \cite{GI85}. However this interpretation faces a fundamental 
problem. The data on the $4\pi$ channels in $e^+e^-$ annihilation are not 
compatible with the $^3P_0$ model \cite{3P0,KI87,ABS96,BCPS97}, which is 
accepted as the most successful model of meson decay. The model works well 
for decays of established ground-state mesons:

$\bullet$ widths predicted to be large, are found to be so

$\bullet$ widths predicted to be small, are found to be so

$\bullet$ calculated widths agree with data to $25 - 40\%$

$\bullet$ signs of amplitudes are correctly predicted

As far as one can ascertain the $^3P_0$ model is reliable, but it has
not been seriously tested for the decays of excited states.

The $^3P_0$ model predicts that the decay of the isovector $2^3S_1$
to $4\pi$ is extremely small:
\be
\Gamma_{2S \to a_1\pi} \sim 3~{\rm MeV}~~~~~~\Gamma_{2S \to h_1\pi} 
\sim 1~{\rm MeV}
\ee
and for the isovector $1^3D_1$ the $a_1\pi$ and $h_1\pi$ decays are large and
equal:
\be
\Gamma_{1D \to a_1\pi} \sim \Gamma_{1D \to h_1\pi} \sim 105~{\rm MeV}
\ee
As $h_1\pi$ contributes only to the $\pi^+\pi^-\pi^0\pi^0$
channel in $e^+e^-$ annihilation, and $a_1\pi$ contributes to both
$\pi^+\pi^-\pi^+\pi^-$ and $\pi^+\pi^-\pi^0\pi^0$, then after subtraction 
of the $\omega\pi$ cross section from the total $\pi^+\pi^-\pi^0\pi^0$
the $^3P_0$ model predicts:
\be
\sigma(e^+e^- \to \pi^+\pi^-\pi^0\pi^0) > \sigma(e^+e^- \to
\pi^+\pi^-\pi^+\pi^-)
\ee
This contradicts observation over much of the available energy range. 
Below $\sim 1.6$ GeV $\sigma(\pi^+\pi^-\pi^+\pi^-) \approx
2\sigma(\pi^+\pi^-\pi^0\pi^0)$, although at higher energies 
$\sigma(\pi^+\pi^-\pi^0\pi^0)$ is the larger.
Further, and more seriously, it has been shown recently by the CMD 
collaboration at Novosibirsk \cite{CMD} and by CLEO \cite{r4a} 
that the dominant channel by far in $4\pi$ (excluding $\omega\pi$) up to 
$\sim 1.6$ GeV is $a_1\pi$. This is quite inexplicable in terms of the $^3P_0$ 
model. So the standard picture is wrong for the isovectors, and there are 
serious inconsistencies in the isoscalar channels as well. One possibility 
is that the $^3P_0$ model is simply failing when applied to excited states, 
which is an intriguing question in itself. An alternative is that there is 
new physics involved. 

\section{A SOLUTION}

A favoured hypothesis is to include vector hybrids \cite{DK93,CP97}, that is 
$q{\bar q}g$ states. The reason for this is that, firstly, hybrid states occur 
naturally in QCD, and secondly, that in the relevant mass range the dominant 
hadronic decay of the isovector vector  hybrid $\rho_H$ is believed to be 
$a_1\pi$ \cite{CP97}. The masses of light-quark hybrids have been obtained in 
lattice-QCD calculations \cite{LMBR97,Ber97,LS98,McN99}, although with quite 
large errors. Results from lattice QCD and other approaches, such as the bag 
model \cite{BC82,CS83}, flux-tube models \cite{IP83}, constituent gluon models 
\cite{KY95} and QCD sum rules \cite{BDY86,LPN87}, show considerable variation 
from each other. So the absolute mass scale is somewhat imprecise, predictions 
for the lightest hybrid lying between 1.3 and 1.9 GeV. However it does seem 
generally agreed that the mass ordering is $0^{-+} < 1^{-+} < 1^{--} < 2^{-+}$.

Evidence for the excitation of gluonic degrees of freedom has emerged in
several processes. Two experiments \cite{E852a,VESa} have evidence for an 
exotic $J^{PC} = 1^{-+}$ resonance, $\hat\rho(1600)$ in the $\rho^0\pi^-$ 
channel in the reaction $\pi^- N \to (\pi^+\pi^-\pi^-) N$. A peak in the 
$\eta\pi$ mass spectrum at $\sim 1400$ MeV with $J^{PC} = 1^{-+}$ in 
$\pi^- N \to (\eta\pi^-) N$ has also been interpreted as a resonance 
\cite{E852b}. Supporting evidence for the 1400 state in the same mode comes 
from ${\bar p}p \to \eta\pi^-\pi^+$ \cite{CB}. There is evidence \cite{VESb} 
for two isovector $0^{-+}$ states in the mass region 1.4 to 1.9 GeV;
$\pi(1600)$ and $\pi(1800)$. The quark model predicts only one. Taking the 
mass of the $1^{-+} \sim 1.4$ GeV, then the $0^{-+}$ is at $\sim 1.3$ GeV 
and the lightest $1^{--}$ at $\sim 1.65$ GeV, which is in the range required 
for the mixing hypothesis to work. Of course if hybrids are comparatively 
heavy, that is the $\hat\rho(1600)$ is the lightest $1^{-+}$ state, and the 
$\pi(1600)$ presumably the corresponding $0^{-+}$ hybrid (or at least with
a significant hybrid component) then the vector hybrid mass $\sim 2.0$ GeV
making strong mixing with the radial and orbital excitations unlikely. 
 
Two specific models for the hadronic hybrids are the flux-tube model 
\cite{CP97,IP83} and the constituent gluon model \cite{Yao85,Kal94}. There 
are some substantial differences in their predictions for hybrid decays. 
For the isovector $1^{--}$ the flux-tube model predicts $a_1\pi$ as 
essentially the only hadronic mode, and a width of $\sim 100$ MeV. The 
constituent gluon model predicts dominant $a_1\pi$, but with significant 
$\rho(\pi\pi)_S$ and $\omega\pi$ components, and a larger width.
For the isoscalar $1^{--}$ the flux-tube model predicts $\rho\pi$ as 
essentially the only hadronic mode, with a width of $\sim 20$ MeV. The
constituent gluon model predicts dominant $\rho\pi$, a significant
$\omega(\pi\pi)_S$ component and a larger width. 

The general conclusion is that the $e^+e^-$ annihilation and 
$\tau$-decay data require the existence of a ``hidden'' vector hybrid in the
isovector and isoscalar channels (assuming that the $^3P_0$ results are
qualitatively reliable). The mixing required is non-trivial, although schemes 
can be devised which are qualitatively compatible with the data \cite{DK99}. 
The unseen physical states are ``off-stage'', in the 1.9 to 2.1 GeV mass 
region. Nonetheless, it appears difficult to achieve quantitative 
agreement with data (within the constraint of specific models) unless the 
hybrids and the $1^3D_1$ states have direct electromagnetic coupling. At 
the simplest level they do not, but these couplings can be generated by 
relativistic corrections at the parton level \cite{GI85} or via intermediate 
hadronic states, for example hybrid $\to$ $a_1\pi$ $\to$ ``$\rho$'' $\to$ 
$e^+e^-$.

To extract the information will require excitation curves for a wide range of 
hadronic final states:
\be
\pi\pi~~~\omega\pi~~~a_1\pi~~~h_1\pi~~~\rho\rho~~~\rho(\pi\pi)_S~~~K{\bar K}~~~
K^*{\bar K}\cdots
\ee
Note that the $n{\bar n}$ states can decay to $K{\bar K}$, $K^*{\bar K}$ etc.
with significant partial widths:
\be
\Gamma_{2S \to K{\bar K}} \sim 30 MeV~~~\Gamma_{1D \to K{\bar K}} \sim 40 MeV
\ee
so isospin separation is necessary in these channels, and there can be mixing
between the isoscalar $n{\bar n}$ states and the $s{\bar s}$ states.

Radiative decays allow a clean separation of the 2S and 1D states.
Preliminary widths in keV are \cite{CDK}:

\begin{center}
\begin{tabular}{|c|c|c|c|c|}
\hline
& $\Gamma(\rho_S)$ & $\Gamma(\omega_S)$ & $\Gamma(\rho_D)$ & 
$\Gamma(\omega_D)$ \\
\hline
$a_1\gamma$ & $\sim 80$ & $\sim 750$ & $\sim 200$ & $\sim 1800$\\
\hline
$a_2\gamma$ & $\sim 100$ & $\sim 900$ & $\sim 10$ & $\sim 100$\\
\hline
$f_1\gamma$ & $\sim 650$ & $\sim 70$ & $\sim 1700$ & $\sim 200$\\
\hline
$f_2\gamma$ & $\sim 1200$ & $\sim 130$ & $\sim 120$ & $\sim 15$\\
\hline
\end{tabular}
\end{center}

Obviously the $f_2\gamma$ channel selects the $\rho_S$ state uniquely.
Additionally, these decays are a much more direct probe of wave functions, 
and hence of models, than are hadronic decay modes.

\section{SUMMARY}

Despite 15 years of work we do not yet understand the light-quark vectors.
Present data raise tantalising questions which go to the heart of 
nonperturbative QCD but are incapable of answering them. These questions 
include:

$\bullet$ How many light-quark vector mesons are there?

$\bullet$ What are their masses, widths, decay channels?

$\bullet$ Do standard models of hadronic decay fail? 

$\bullet$ What hybrid states are hiding in there?

$\bullet$ What is the nature of hybrids: flux tube or constituent gluon?

$\bullet$ Are the masses of hybrids compatible with lattice QCD?

High-statistics, comprehensive $e^+e^-$ annihilation data provides by far 
the best way to answer these and related questions. The data are a unique
$J^{PC}=1^{--}$ laboratory. 

Whatever the answers, new physics is guaranteed!


\begin{thebibliography}{99}

\bibitem{EO86}
C. Erkal and M.G. Olsson: Z.Phys. {\bf C31} (1986) 615

\bibitem{DM87}
A. Donnachie and H. Mirzaie: Z.Phys. {\bf C33} (1987) 407

\bibitem{PDG}
Particle Data Group: European Physical Journal C15 (2000) 1

\bibitem{r1} 
L. M. Barkov et al: Nucl.Phys. {\bf B256} (1985) 365\\
D. Bisello et al: Phys.Lett. {\bf B220} (1989) 321

\bibitem{r2} 
S. I. Dolinsky et al: Phys.Lett. {\bf B174}, (1986) 453

\bibitem{r3} 
R. Barate et al (ALEPH Collaboration): Z.Phys. {\bf C76} (1997) 15

\bibitem{r4} 
H. Albrecht et al (ARGUS Collaboration): Phys.Lett. {\bf B185} (1987) 223

\bibitem{r4a}
K.W. Edwards et al (CLEO Collaboration): Phys.Rev {\bf D61} (2000) 072003

\bibitem{r4b}
S. Anderson et al (CLEO Collaboration): Phys.Rev {\bf D61} (2000) 112002

\bibitem{r5} 
S. I. Dolinsky et al: Phys.Rep. {\bf 202} (1991) 99

\bibitem{r5a}
L Stanco (DM2 Collaboration): Proc. Hadron'91, Maryland, 1991; ed Y. Oneda
and D. Peaslee (World Scientific, Singapore, 1992) p.84

\bibitem{r6} 
C. Bacci et al: Nucl.Phys. {\bf B184} (1981) 31\\
G. Cosme et al: Nucl.Phys. {\bf B152} (1979) 215

\bibitem{r7} 
A. B. Clegg and A. Donnachie: Z.Phys. {\bf C62} (1994) 455\\
A. Donnachie and A.B. Clegg: Phys.Rev. {\bf D51} (1995) 4979

\bibitem{r8} 
N.N. Achasov and A.A. Kozhevnikov: Phys.Rev. {\bf D55} (1997) 2663;
Phys.Rev. {\bf D62} (2000) 117503

\bibitem{r9} 
A. Antonelli et al: Phys.Lett. {\bf B212} (1988) 133

\bibitem{r10} 
S. Fukui et al: Phys.Lett. {\bf B202} (1988) 133

\bibitem{r11} 
A. Abele et al (Crystal Barrel Collaboration): Phys.Lett. {\bf B391} (1997) 191

\bibitem{r12} 
A. Antonelli et al: Z.Phys. {\bf 56} (1992) 15

\bibitem{SND}
M.N. Achasov et al (SND Collaboration): Phys.Lett. {\bf B462} (1999) 365

\bibitem{GI85}
S. Godfrey and N. Isgur: Phys.Rev. {\bf D32} (1985) 189

\bibitem{3P0} 
G. Busetto and L. Oliver: Z.Phys. {\bf C20} (1983) 247\\
P. Geiger and E.S. Swanson: Phys.Rev. {\bf D50} (1994) 6855\\
H.G. Blundell and S. Godfrey: Phys.Rev. {\bf D53} (1996) 3700

\bibitem{KI87} 
R. Kokoski and N. Isgur: Phys.Rev. {\bf D35} (1987) 907

\bibitem{ABS96} 
E.S. Ackleh, T. Barnes and E.S. Swanson: Phys.Rev. {\bf D54} (1996) 6811

\bibitem{BCPS97} 
T. Barnes, F.E. Close, P.R. Page and E.S. Swanson: Phys.Rev. {\bf D55} (1997)
4157

\bibitem{CMD}
R.R. Akhmetshin et al (CMD Collaboration): Phys.Lett. {\bf B466} (1999) 392

\bibitem{DK93}
A. Donnachie and Yu.S. Kalashnikova: Z.Phys. {\bf C59} (1993) 621

\bibitem{CP97}
F.E. Close and P.R. Page: Phys.Rev. {\bf D56} (1997) 1584

\bibitem{LMBR97} 
P. Lacock, C. Michael, P. Boyle and P. Rowland: Phys.Lett. {\bf B401} 
(1997) 308 

\bibitem{Ber97}
C. Bernard et al: Phys.Rev {\bf D56} (1997) 7039; Nucl.Phys.Proc.Supp. 73
(1999) 264

\bibitem{LS98}
P. Lacock and K. Schilling: Nucl.Phys.Proc.Supp. 73 (1999) 261

\bibitem{McN99}
C. McNeile: hep-lat/9904013

\bibitem{BC82} 
T. Barnes and F.E. Close: Phys.Lett. {\bf 116B} (1982) 365; 
{\it ibid} {\bf 123B} (1983) 89\\
T. Barnes, F.E. Close and F. de Viron: Nucl.Phys. {\bf B224} (1983) 241

\bibitem{CS83}
M. Chanowitz and S. Sharpe: Nucl.Phys. {\bf B222} (1983) 211

\bibitem{IP83}
N. Isgur and J.E. Paton: Phys.Lett. {\bf 124B} (1983) 247\\
N. Isgur, R. Kokoski and J.E. Paton: Phys.Rev.Lett. {\bf 54}, 869 
(1985)\\
N. Isgur and J.E. Paton: Phys.Rev. {\bf D31} (1985) 2910\\
T. Barnes, F.E. Close, E.S. Swanson: Phys.Rev. {\bf D52} (1995) 5242

\bibitem{KY95}
Yu.S. Kalashnikova and Yu.B. Yufryakov: Phys. Lett. {\bf B359} (1995) 175\\
Yu.S. Kalashnikova and Yu.B.Yufryakov: Phys. At. Nucl. {\bf 60} (1997) 307  

\bibitem{BDY86}
I.I. Balitsky, D.I. Dyakonov and A.V. Yung: Z.Phys. {\bf C33} (1986) 265

\bibitem{LPN87}
J.I. Latorre, P. Pascual and S. Narison: Z.Phys. {\bf C34} (1987) 347

\bibitem{E852a} 
D.P. Weygand (E852 Collaboration): Proc. HADRON'97, BNL; ed S-U Chung and 
H.J. Willutski (American Institute of Physics, New York, 1998) p.313

\bibitem{VESa} 
Yu.P. Gouz (VES Collaboration): Proc. XXVI ICHEP (DALLAS, 1992),ed. 
J.R. Sanford, p.572

\bibitem{E852b} 
D.R. Thompson et al (E852 Collaboration): Phys.Rev.Lett. {\bf 79} (1997) 1630

\bibitem{CB} 
A. Abele et al (Crystal Barrel Collaboration): Phys.Lett. {\bf B423} (1998) 175

\bibitem{VESb} 
A. Zaitsev (VES Collaboration): Proc. Hadron'97, BNL, 1997; ed S-U Chung and 
H.J. Willutski (American Institute of Physics, New York, (1998) p.461\\
D V Amelin (VES Collaboration): {\it ibid} p.770

\bibitem{Yao85}
A. Le Yaounac, L. Oliver, O. P\`ene, J.C. Raynal and S. Ono: Z.Phys.
{\bf C28} (1985) 309\\
F. Iddir, A. Le Yaouanc, L. Oliver, O. P\`ene and J.C. Raynal:
Phys.Lett. {\bf B205} (1988) 564

\bibitem{Kal94}
Yu.S. Kalashnikova: Z.Phys. {\bf C62} (1994) 323

\bibitem{DK99}
A. Donnachie and Yu.S. Kalashnikova: Phys.Rev. {\bf D60} (1999) 114011 

\bibitem{CDK} 
F.E. Close, A. Donnachie and Yu.S. Kalashnikova: in preparation







\end{thebibliography}
\end{document}